\documentclass{ws-ijmpa}
\usepackage[super,compress]{cite}
\usepackage{graphicx}
\usepackage{leftidx}
\usepackage{caption2}
\begin{document}
	\markboth{CHANDNI MENAPARA and AJAY KUMAR RAI}{Spectra of $\Lambda$ and $\Sigma$ Baryons under Screened Potential }
	
	%
	\catchline{}{}{}{}{}
	%
	
	\title{Spectra of $\Lambda$ and $\Sigma$ Baryons under Screened Potential }

	\author{CHANDNI MENAPARA\footnote{chandni.menapara@gmail.com}     and AJAY KUMAR RAI}
	
	\address{Department of Physics, The Maharaja Sayajirao University of Baroda, Vadodara, India\\
		Department of Physics, Sardar Valabhbhai National Institute of Technology\\
	Surat-395008, Gujarat, India
	\\
		}
	
%
	
	\maketitle
	
	\begin{history}
		\received{Day Month Year}
		\revised{Day Month Year}
	\end{history}
	
	\begin{abstract}
	The light, strange baryons have been studied through various approaches and attempted to be looked for rigorously in experiments. {\bf The screened potential has been applied to heavy baryon sector as well as meson systems in earlier works. Here, this article attempts to compare the results for linear and screened potential for light strange baryons.} Also, the Regge trajectories depict the linear nature. 
		
	\end{abstract}
	
	\ccode{PACS numbers:}
	
\section{Introduction}
\label{sec1}
JLab, LHC, BESIII experimental facilities have over the years looked for missing resonances from light to heavy quark systems \cite{lhc,bes}. An upcoming experiment PANDA is highly focused to look for strange baryon especially ones with higher strangeness $\Xi$ and $\Omega$ \cite{panda2,panda3,panda4,panda5,panda6,panda7}. With every run and new upgrades in energy scale, exotic states have come up however strange quark still hasn't revealed all its composites easily. Singly strange $\Lambda$ and its similar partner $\Sigma$ have still a sizable number of resonances but many are 1 and 2 star status \cite{pdg}. A non-relativistic approach under constituent cover with different confining potential attempts to reproduce these excited states. This is the driving force for the Hadron spectroscopy field \cite{thiel, eichmann}. \\
 
 Many approaches have been used throughout the years in an effort to fully comprehend the baryon sector. Internal baryon dynamics have been studied using algebraic models; a recent study used U(7) in this regard \cite{amiri}. R. Bijker and group too have focused on algebraic approach with the string-like model \cite{bijker}.  the quark-diquark model has been examined in several variations; E. Santopinto et al. used an exchange interaction inspired by Gursey Radicati to generate both strange and non-strange baryon resonances \cite{s5,s15}. Regge phenomenology has also been employed in the study of light, strange baryons using n and J plane linear curves \cite{juhi}. Other than these various models have been employed over the years \cite{chen,klempt}.\\
\noindent The screened potential term is accompanied with spin-dependent term in the present approach as detailed in section 2. The section 3 discusses mass spectra of $\Lambda$, $\Sigma$ baryons along with the experimental and theoretical background known so far Few states need special attention regarding their actual nature. Section 4 is dedicated to Regge trajectory for (J,$M^2$) and (n,$M^2$).

\section{Theoretical Framework}
\label{sec2}
Hypercentral Constituent Quark Model (hCQM) has been applied in various systems and with a variety of potentials for heavy hadrons \cite{zalak19,zalak18,zalak16,zalak17,keval18,keval20,amee1,amee2,amee3} and exotics. The linear potential has been employed for all octet, decuplet baryons in our earlier works \cite{cpc1,cpc2,cpc3,ijmpa1,ijmpa2,akram}. The screened potential now applied to strange $\Lambda$ and $\Sigma$ baryons, allows to give a vivid comparison between linear and screened potential. Our earlier works have highlighted the detailed comparison with other theoretical models. \\
Employing the Jacobi coordinates to describe the three body baryon system as marked by notable articles \cite{morpurgo, simonov, ballot},

\begin{equation}
	\vec{\rho}=\frac{\vec{r_1}-\vec{r_2}}{\sqrt2} \hspace{0.5cm} and \hspace{0.5cm} 
	\vec{\lambda}=\frac{\vec{r_1}+\vec{r_2}-2\vec{r_3}} {\sqrt{6}}.
	\label{1}
\end{equation} 

\noindent Hyperradias $x$ and hyperangle $\xi$ in terms of Jacobi coordinates are \cite{gianinni},

\begin{equation}
	x=\sqrt{\rho^2+\lambda^2} 
	\hspace{0.5cm} and \hspace{0.5cm}
	\xi=arctan\left(\frac{\rho}{\lambda}\right)
	\label{2}
\end{equation}

\noindent The Hamiltonian, presenting the three quark bound system is,

\begin{equation} 
	H=\frac{P^2}{2m} + V(x)
	\label{3}
\end{equation}

\noindent Here, $P$ is conjugate momentum and $m$ is the reduced mass of the system, which expressed as, $m=\frac{2m_\rho m_\lambda}{m_\rho + m_\lambda}$. Here, the constituent quark mass $m_u=m_d=0.290 GeV$, $m_s=0.500 GeV$. {\bf As the hypercentral model itself suggests, $V(x)$ is non-relativistic interaction potential inside the baryonic system  depending only on hyperradius x \cite{gianinni, giannini1, giannini2}.}

\noindent The screened potential is incorporated as confining potential with the color-Coulomb potential (spin independent  potential $V_{SI}(x)=V_{conf}(x) + V_{Col}(x)$). 

\begin{equation}
	V_{conf}(x)=a\left(\frac{1-e^{-{\mu} x}}{\mu}\right)  \hspace{0.5cm}  and \hspace{0.5cm} 	V_{Col}(x)= -\frac{2}{3}\frac{\alpha_s}{x}
	\label{5}
\end{equation}

\noindent where, $a$ is the string tension. {\bf Based on a paper by R. Chaturvedi, the screening parameter $\mu$ has been varied over a range and 0.3 has been considered as the value obtain the spectra for all the systems considered here \cite{raghav} for the case of mesons. }
The spin dependent part of potential $V_{SD}(x)$ is,
\begin{equation}
	V_{SD}(x) = V_{SS}(x)(\vec{S_{\rho}} \cdot \vec{S_{\lambda}}) + V_{\gamma S}(x)(\vec{\gamma} \cdot \vec{S}) +V_T(x) \left[ S^2 - \frac{3 (\vec{S} \cdot \vec{x}) (\vec{S} \cdot \vec{x})}{x^2} \right]
\end{equation}
which includes spin-spin, spin-orbit and tensor term respectively. 
{\bf In case of linear potential for the similar articles mentioned above, the differences for the resonances predicted for higher excited hyperfine states were more. The screened effect has shown effects in heavy baryons by reducing this splitting, which has been tried to observe in higher excited light baryons.}
\section{Mass Spectra}
\label{sec3}
The PDG signifies the currently known resonances of $\Lambda$ with 10 four star states, 4 three star states, 2 two star states and 7 one star states. Except for four star states, there is a wide opportunity to look for the resonances in every possible reactions to know all the details. The very first excited state 1405 MeV has been under the spotlight for years. In some instances, the measurement results are consistent within a relatively narrow range. This is true not only for the well-known (1520)$\frac{3}{2}$, but also for the (1670)$\frac{1}{2}$, (1690)$\frac{3}{2}$, and (1815)$\frac{5}{2}^{+}$, all of which lie within a relatively narrow mass range.
The findings of (1890) $\frac{3}{2}^{+}$, (1830)$\frac{5}{2}$, and (2100)$\frac{7}{2}$ are also very compelling. A Four star status has been assigned to these resonances. As it is evident that the low-lying masses i.e. S-wave the predicted mass differ by 50 MeV from PDG. Also, lower excited states for P and D-wave also, the masses are quite in the range with difference of 20-30 MeV. But for 1F, 1G the masses are under-predicted by more than 100 MeV to experimental as well as that of linear potential ones. \\

$\Sigma$ baryon happens to have a large number of states with 1 and 2 stars clearly marking the need to look for resonances. In the earliest studies, the resonances (1915)$\frac{5}{2}^{+}$ and (1910)$\frac{3}{2}^{-}$  emerge as the leading candidates, with additional support for (1880) $\frac{1}{2}^{+}$ and (1900)$\frac{1}{2}^{+}$. The four star states (1670)$\frac{3}{2}^{-}$, (1775)$\frac{5}{2}^{-}$, (1915)$\frac{5}{2}^{+}$, and (2030)$\frac{7}{2}^{+}$ have been established with good consistency. As here (1915)$\frac{5}{2}^{+}$ and (2030)$\frac{7}{2}^{+}$ are both assigned to 1D family, our masses are higher by 37 MeV compared to 1915. State 1660 is 2S($\frac{1}{2}^{+}$) differs by 30 MeV. The three star state 2230 doesn't have known spin-parity but here we have tentatively assigned it to be 3S($\frac{3}{2}^{+}$).

\begin{table}
	
	\caption{$\Lambda$ Resonance mass spectra using Screened potential (in MeV)}
	\label{tab:lambda-screen} 
		\renewcommand{\arraystretch}{1.5}
		\centering
		\begin{tabular}{cccccccccccc}
			\hline
			State & $J^{P}$ & $M_{scr}$ & $M_{lin}$ & $M_{exp}$\cite{pdg} & \cite{faustov} & \cite{bijker1} & \cite{melde} & \cite{chen1} \\
			\hline
			1S & $\frac{1}{2}^{+}$ & 1115 & 1115 & 1115 &  1115 & 1133 & 1136 & 1113 \\
			2S & $\frac{1}{2}^{+}$ & 1553 & 1589 & 1600 & 1615 & 1577 & 1625 & 1606\\
			3S & $\frac{1}{2}^{+}$ & 1869 & 1892 & 1810 & 1901 & & & 1880 \\
			4S & $\frac{1}{2}^{+}$ & 2197 & 2220 & & 1986 & & & 2173\\
			5S & $\frac{1}{2}^{+}$ & 2536 & 2571 & & 2099 & & & \\
			\hline
			$1^{2}P_{1/2}$ & $\frac{1}{2}^{-}$ & 1554 & 1558 & & 1667 & 1686 & 1556 & 1559 \\
			$1^{2}P_{3/2}$ & $\frac{3}{2}^{-}$ & 1549 & 1544 & 1520 & 1549 & & & 1560 \\
			$1^{4}P_{1/2}$ & $\frac{1}{2}^{-}$ & 1557 & 1564 & 1670 & & & & 1656 \\
			$1^{4}P_{3/2}$ & $\frac{3}{2}^{-}$ & 1552 & 1551 & 1690 & 1693 & & &\\
			$1^{4}P_{5/2}$ & $\frac{5}{2}^{-}$ & 1545 & 1533 & & & & & \\
			\hline
			$2^{2}P_{1/2}$ & $\frac{1}{2}^{-}$ & 1814 & 1858 &  1800 & 1733 & & & 1791 \\
			$2^{2}P_{3/2}$ & $\frac{3}{2}^{-}$ & 1808 & 1841 & & 1812 & & & 1859 \\
			$2^{4}P_{1/2}$ & $\frac{1}{2}^{-}$ & 1817 & 1867 & & &\\
			$2^{4}P_{3/2}$ & $\frac{3}{2}^{-}$ & 1811 & 1850 & & &\\
			$2^{4}P_{5/2}$ & $\frac{5}{2}^{-}$ & 1803 & 1827 & 1830 & 1861 & 1799 & 1778 & 1803 \\
			\hline
			$3^{2}P_{1/2}$ & $\frac{1}{2}^{-}$ & 2083 & 2186 & & 2155 & \\
			$3^{2}P_{3/2}$ & $\frac{3}{2}^{-}$ & 2077 & 2166 & & 2035 & \\
			$3^{4}P_{1/2}$ & $\frac{1}{2}^{-}$ & 2086 & 2196 & & 2197 & \\
			$3^{4}P_{3/2}$ & $\frac{3}{2}^{-}$ & 2080 & 2176 & & & \\
			$3^{4}P_{5/2}$ & $\frac{5}{2}^{-}$ & 2072 & 2149 & & 2136 & \\
			\hline
			$1^{2}D_{3/2}$ & $\frac{3}{2}^{+}$ & 1765 & 1789 & & 1854 & 1849 & & 1836 \\
			$1^{2}D_{5/2}$ & $\frac{5}{2}^{+}$ & 1755 & 1767& 1820 & 1825 & 1849 & & 1839 \\
			$1^{4}D_{1/2}$ & $\frac{1}{2}^{+}$ & 1776 & 1814 &1710* & 1901 & 1799 & 1799 & 1764 \\
			$1^{4}D_{3/2}$ & $\frac{3}{2}^{+}$ & 1768 & 1798 & 1890 & & & & & \\
			$1^{4}D_{5/2}$ & $\frac{5}{2}^{+}$ & 1758 & 1776 & & & & & \\
			$1^{4}D_{7/2}$ & $\frac{7}{2}^{+}$ & 1746 & 1748 & & & & & \\
			\hline
			$2^{2}D_{3/2}$ & $\frac{3}{2}^{+}$ & 2032 & 2113 & 2070 \\
			$2^{2}D_{5/2}$ & $\frac{5}{2}^{+}$ & 2022 & 2085 & 2110 \\
			\hline
		\end{tabular}
	\end{table}

\begin{table}
		\renewcommand{\arraystretch}{1.5}
		\centering
		\begin{tabular}{ccccccccccc}
			\hline
			State & $J^{P}$ & $M_{scr}$ & $M_{lin}$ & $M_{exp}$ \cite{pdg}  & \cite{faustov} & \cite{bijker1} & \cite{melde} & \cite{chen1}\\
			\hline
			$2^{4}D_{1/2}$ & $\frac{1}{2}^{+}$ & 2043 & 2144 & & & \\
			$2^{4}D_{3/2}$ & $\frac{3}{2}^{+}$ & 2036 & 2123 & & & \\
			$2^{4}D_{5/2}$ & $\frac{5}{2}^{+}$ & 2026 & 2096 & & & 2074 & & 2008 \\
			$2^{4}D_{7/2}$ & $\frac{7}{2}^{+}$ & 2013 & 2061 & 2085 & & & & 2064\\
			\hline
			$3^{2}D_{3/2}$ & $\frac{3}{2}^{+}$ & 2306 & 2459 & & & \\
			$3^{2}D_{5/2}$ & $\frac{5}{2}^{+}$ & 2296 & 2426 & & & \\
			$3^{4}D_{1/2}$ & $\frac{1}{2}^{+}$ & 2316 & 2496 & & & \\
			$3^{4}D_{3/2}$ & $\frac{3}{2}^{+}$ & 2309 & 2471 & & & \\
			$3^{4}D_{5/2}$ & $\frac{5}{2}^{+}$ & 2300 & 2438 & & & \\
			$3^{4}D_{7/2}$ & $\frac{7}{2}^{+}$ & 2288 & 2398 & & & \\
			\hline
			
			$1^{2}F_{5/2}$ & $\frac{5}{2}^{-}$ & 1980 & 2039 &2080 & & \\
			$1^{2}F_{7/2}$ & $\frac{7}{2}^{-}$ & 1965 & 2002 & 2100 & 2097 & \\
			$1^{4}F_{3/2}$ & $\frac{3}{2}^{-}$ & 1996 & 2079 & & & \\
			$1^{4}F_{5/2}$ & $\frac{5}{2}^{-}$ & 1984 & 2050 & & & \\
			$1^{4}F_{7/2}$ & $\frac{7}{2}^{-}$ & 1969 & 2013 & & & \\
			$1^{4}F_{9/2}$ & $\frac{9}{2}^{-}$ & 1951 & 1969 & & \\
			\hline
			$2^{2}F_{5/2}$ & $\frac{5}{2}^{-}$ & 2252 & 2380 \\
			$2^{2}F_{7/2}$ & $\frac{7}{2}^{-}$ & 2239 & 2337 \\
			$2^{4}F_{3/2}$ & $\frac{3}{2}^{-}$ & 2267 & 2427 & 2325 \\
			$2^{4}F_{5/2}$ & $\frac{5}{2}^{-}$ & 2256 & 2393 \\
			$2^{4}F_{7/2}$ & $\frac{7}{2}^{-}$ & 2243 & 2350 \\
			$2^{4}F_{9/2}$ & $\frac{9}{2}^{-}$ & 2226 & 2299 \\
			\hline
			$1^{2}G_{7/2}$ & $\frac{7}{2}^{+}$ & 2198 & 2302 & & 2251 & & & \\
			$1^{2}G_{9/2}$ & $\frac{9}{2}^{+}$ & 2179 & 2246 & 2350 & 2360 & 2357 & & \\
			$1^{4}G_{5/2}$ & $\frac{5}{2}^{+}$ & 2219 & 2363 & & 2258 & & & \\
			$1^{4}G_{7/2}$ & $\frac{7}{2}^{+}$ & 2203 & 2316 & & & & & \\
			$1^{4}G_{9/2}$ & $\frac{9}{2}^{+}$ & 2184 & 2260 & & & & & \\
			$1^{4}G_{11/2}$ & $\frac{11}{2}^{+}$ & 2162 & 2195 & & & & & \\
			\hline
		\end{tabular}
	
\end{table}

\begin{table}
	\caption{$\Sigma$ resonance spectra using Screened potential (in MeV)}
	\label{tab:Sigma-screen}

		\centering
		\begin{tabular}{cccccccccc}
			\hline
			State & $J^{P}$ & $M_{scr}$ & $M_{lin}$ & $M_{exp}$\cite{pdg} & \cite{faustov} & \cite{bijker1} & \cite{melde} & \cite{chen1}\\
			\hline
			1S & $\frac{1}{2}^{+}$ & 1193 & 1193 & 1193 & 1187 & 1170 & 1180 & 1192\\
			& $\frac{3}{2}^{+}$ & 1382 & 1384 & 1385 & 1381 & 1832 & 1389 & 1383 \\
			2S & $\frac{1}{2}^{+}$ & 1630 & 1643 &1660 & 1711 & 1604 & 1616 & 1664 \\
			& $\frac{3}{2}^{+}$ & 1809 & 1827 & 1780 & 1862 & & 1865 & 1868\\
			3S & $\frac{1}{2}^{+}$ & 2035 & 2099 & & 2028 & & & 2022\\
			& $\frac{3}{2}^{+}$  & 2186 & 2236 & 2230 \\
			4S & $\frac{1}{2}^{+}$ & 2464 & 2589 \\
			& $\frac{3}{2}^{+}$ & 2597 & 2693 \\ 
			5S & $\frac{1}{2}^{+}$ & 2916 & 3108\\
			& $\frac{3}{2}^{+}$ & 3034 & 3189 \\
			\hline
			$1^{2}P_{1/2}$ & $\frac{1}{2}^{-}$ & 1679 & 1725 & 1620 & 1620 & 1711 & 1677 & 1657 \\
			$1^{2}P_{3/2}$ & $\frac{3}{2}^{-}$ & 1669 & 1702 & 1670 & 1706 & 1711 & 1677 & 1698 \\
			$1^{4}P_{1/2}$ & $\frac{1}{2}^{-}$ & 1683 & 1736 &  1750 & 1693 & & 1736 & 1746 \\
			$1^{4}P_{3/2}$ & $\frac{3}{2}^{-}$ & 1674 & 1713 & & 1731 & & 1736 & 1790 \\
			$1^{4}P_{5/2}$ & $\frac{5}{2}^{-}$ & 1662 & 1683 & 1775 & 1757 & & 1736 & 1743 \\
			\hline
			$2^{2}P_{1/2}$ & $\frac{1}{2}^{-}$ & 2036 & 2145 &1900 & 2115 & 2110 & &\\
			$2^{2}P_{3/2}$ & $\frac{3}{2}^{-}$ & 2025 & 2114 &1910 & 2175 & & & \\
			$2^{4}P_{1/2}$ & $\frac{1}{2}^{-}$ & 2042 & 2159 &2110\\
			$2^{4}P_{3/2}$ & $\frac{3}{2}^{-}$ & 2030 & 2129 &2010\\
			$2^{4}P_{5/2}$ & $\frac{5}{2}^{-}$ & 2015 & 2087 & \\
			\hline
			$3^{2}P_{1/2}$ & $\frac{1}{2}^{-}$ & 2417 & 2608\\
			$3^{2}P_{3/2}$ & $\frac{3}{2}^{-}$ & 2406 & 2571 \\
			$3^{4}P_{1/2}$ & $\frac{1}{2}^{-}$ & 2422 & 2627\\
			$3^{4}P_{3/2}$ & $\frac{3}{2}^{-}$ & 2411 & 2589\\
			$3^{4}P_{5/2}$ & $\frac{5}{2}^{-}$ & 2396 & 2541\\
			\hline
			$1^{2}D_{3/2}$ & $\frac{3}{2}^{+}$ & 1971 & 2057 & 1940 & 2025 & & & 1947 \\
			$1^{2}D_{5/2}$ & $\frac{5}{2}^{+}$ & 1952 & 2013 & 1915 & 1991 & 1872 & & 1949\\
			\hline
		\end{tabular}
	\end{table}
	\begin{table}
		\centering
		\begin{tabular}{ccccccccc}
			\hline
			State & $J^{P}$ & $M_{scr}$ & $M_{lin}$ &$M_{exp}$\cite{pdg}  & \cite{faustov} & \cite{bijker1} & \cite{melde} & \cite{chen1}\\
			\hline
			
			$1^{4}D_{1/2}$ & $\frac{1}{2}^{+}$ & 1991 & 2107 & 1983 & & 1911 & 1924 \\
			$1^{4}D_{3/2}$ & $\frac{3}{2}^{+}$ & 1978 & 2074\\
			$1^{4}D_{5/2}$ & $\frac{5}{2}^{+}$ & 1959 & 2029\\
			$1^{4}D_{7/2}$ & $\frac{7}{2}^{+}$ & 1937 & 1974 &2025 & 2033 & & & 2002 \\
			\hline
			$2^{2}D_{3/2}$ & $\frac{3}{2}^{+}$ & 2347 & 2510\\
			$2^{2}D_{5/2}$ & $\frac{5}{2}^{+}$ & 2329 & 2459\\
			$2^{4}D_{1/2}$ & $\frac{1}{2}^{+}$ & 2368 & 2568\\
			$2^{4}D_{3/2}$ & $\frac{3}{2}^{+}$ & 2354 & 2529\\
			$2^{4}D_{5/2}$ & $\frac{5}{2}^{+}$ & 2336 & 2478\\
			$2^{4}D_{7/2}$ & $\frac{7}{2}^{+}$ & 2313 & 2414 & 2470 \\
			\hline
			$3^{2}D_{3/2}$ & $\frac{3}{2}^{+}$ & 2747 & 3004 & 3000*\\
			$3^{2}D_{5/2}$ & $\frac{5}{2}^{+}$ & 2727 & 2945\\
			$3^{4}D_{1/2}$ & $\frac{1}{2}^{+}$ & 2769 & 3072\\
			$3^{4}D_{3/2}$ & $\frac{3}{2}^{+}$ & 2754 & 3027\\
			$3^{4}D_{5/2}$ & $\frac{5}{2}^{+}$ & 2735 & 2967\\
			$3^{4}D_{7/2}$ & $\frac{7}{2}^{+}$ & 2710 & 2892\\
			\hline
			$1^{2}F_{5/2}$ & $\frac{5}{2}^{-}$ & 2279 & 2416\\
			$1^{2}F_{7/2}$ & $\frac{7}{2}^{-}$ & 2249 & 2343 & 2100*\\
			$1^{4}F_{3/2}$ & $\frac{3}{2}^{-}$ & 2311 & 2495 & & 2300 & & & \\
			$1^{4}F_{5/2}$ & $\frac{5}{2}^{-}$ & 2287 & 2437 & & 2347 & & & \\
			$1^{4}F_{7/2}$ & $\frac{7}{2}^{-}$ & 2258 & 2365 & & 2289 & & & \\
			$1^{4}F_{9/2}$ & $\frac{9}{2}^{-}$ & 2223 & 2278 & & 2289 & & & \\
			\hline
			$2^{2}F_{5/2}$ & $\frac{5}{2}^{-}$ & 2681 & 2901 \\
			$2^{2}F_{7/2}$ & $\frac{7}{2}^{-}$ & 2653 & 2819 \\
			$2^{4}F_{3/2}$ & $\frac{3}{2}^{-}$ & 2704 & 2990 \\
			$2^{4}F_{5/2}$ & $\frac{5}{2}^{-}$ & 2681 & 2925 \\
			$2^{4}F_{7/2}$ & $\frac{7}{2}^{-}$ & 2653 & 2844 \\
			$2^{4}F_{9/2}$ & $\frac{9}{2}^{-}$ & 2619 & 2746 \\
			\hline
		\end{tabular}
\end{table}
The low-lying states are matching but higher excited states are under-predicted. However, the mass-range suggest that the given spin-parity assignment to one and two star states are well in accordance. From the resonance tables \ref{tab:lambda-screen} and \ref{tab:Sigma-screen}, it is noteworthy that for states with higher angular momentum, the under-predicted masses show the effect of screened potential. Also, the hypercentral potential leads to reversed hierarchy for hyperfine states. \\
The $M_{lin}$ shown in tables 1 and 2 gives the resonance masses predicted through linear potential in our earlier works for the comparison with screened potential results. It is noteworthy that linear masses have greater difference in the hyperfine states as in case of screened potential. Thus, at very high orbital states linear potential gives over prediction for the masses. This serves as a primary aim to observe the difference in application of both these potentials. Also, the tables above showcases comparison with few of the models described in section 1. The relativistic quark model studies by Faustov et al \cite{faustov} have been incorporated which shows that few states predicted through screened potential are quite in accordance. 

\section{Regge Trajectory}
\begin{figure}
	\centering
	\includegraphics[scale=0.3]{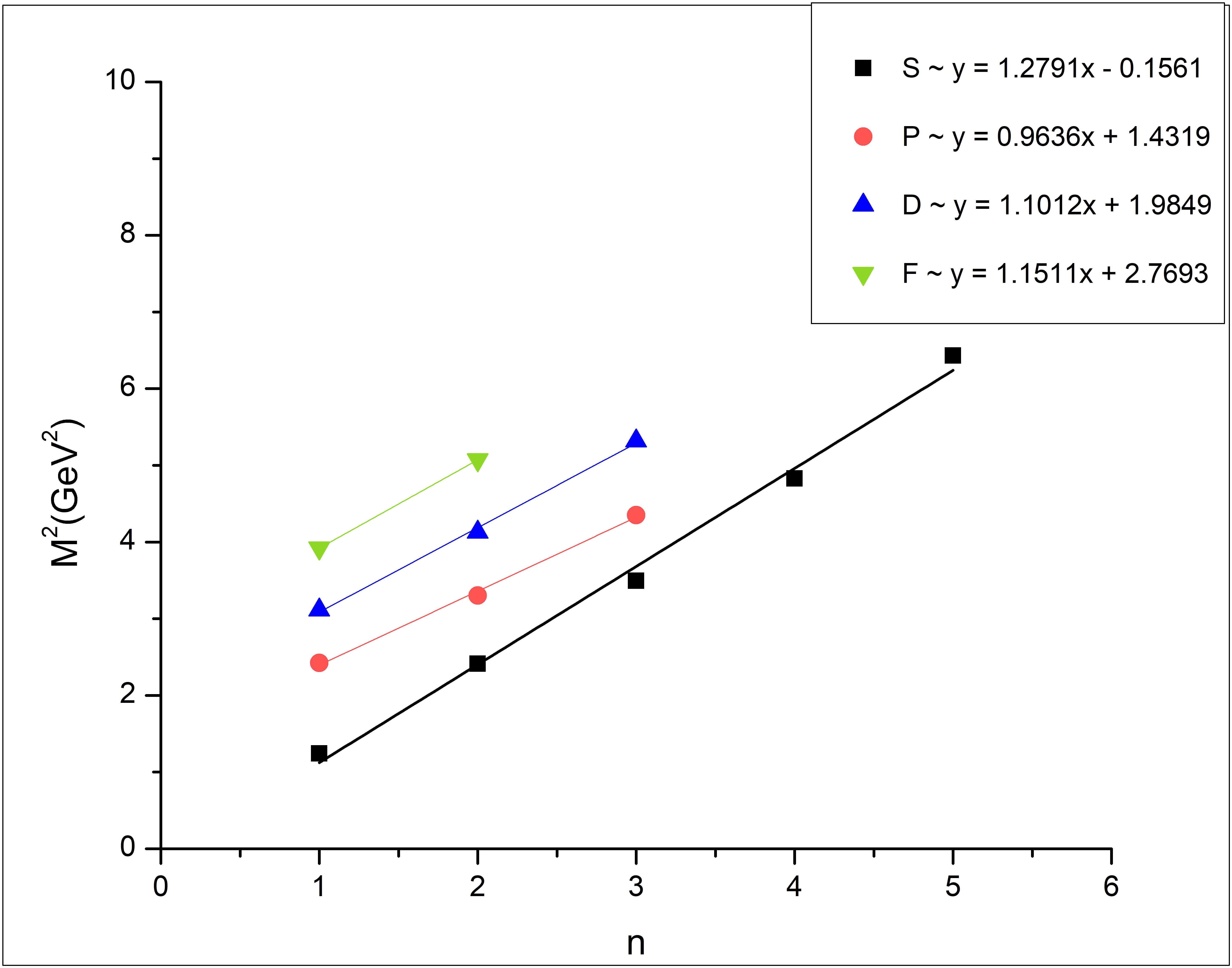}
	\caption{$n$ vs $M^{2}$ for $\Lambda$, Regge trajectory for Principle Quantum number n versus $M^{2}$ depicting the linear nature. }
	\label{fig:lscr}
\end{figure}

\begin{figure}
	\centering
	\includegraphics[scale=0.3]{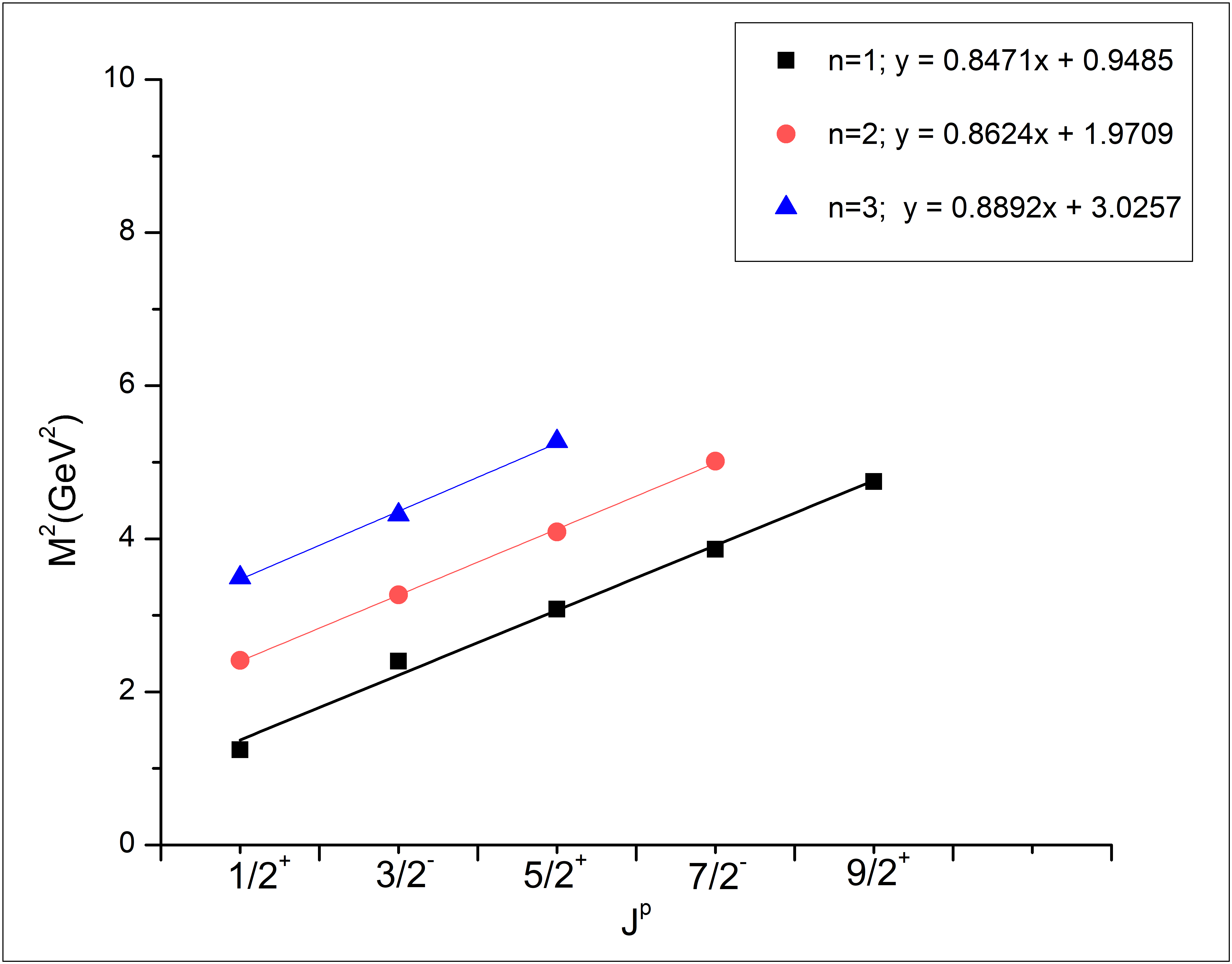}
	\caption{$J^P$ vs $M^{2}$ for $\Lambda$, Regge trajectory for Angular Momentum Quantum number J versus $M^{2}$ for natural parity.}
	\label{fig:ljscr}
\end{figure}

\begin{figure}
	\centering
	\includegraphics[scale=0.3]{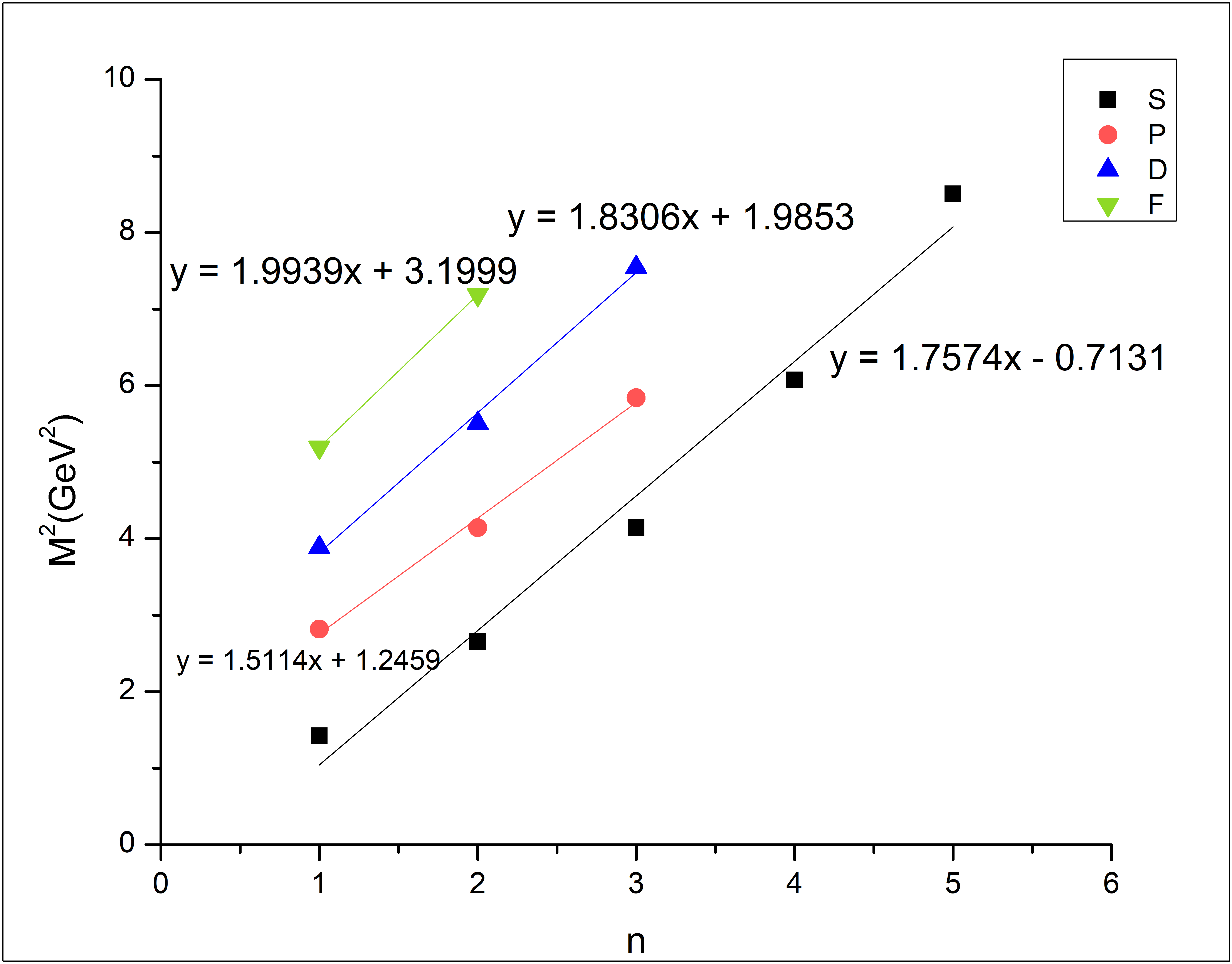}
	\caption{$n$ vs $M^{2}$ for $\Sigma$, Regge trajectory for Angular Momentum Quantum number J versus $M^{2}$.}
	\label{fig:sscr}
\end{figure}

\begin{figure}
	\centering
	\includegraphics[scale=0.3]{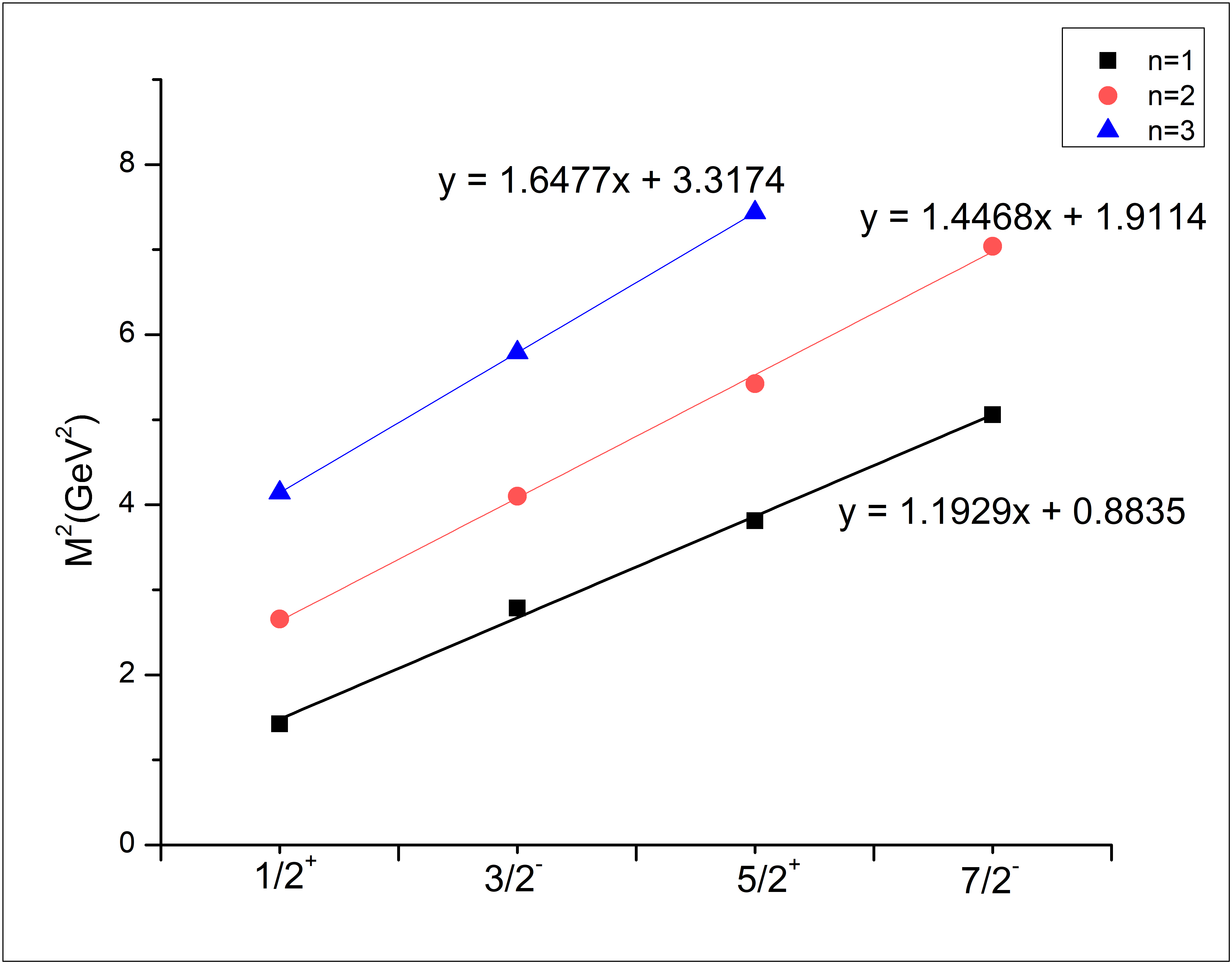}
	\caption{$J^P$ vs $M^{2}$ for $\Sigma$, Regge trajectory for Angular Momentum Quantum number J versus $M^{2}$ for natural parity.}
	\label{fig:sj1scr}
\end{figure}

\begin{figure}
	\centering
	\includegraphics[scale=0.3]{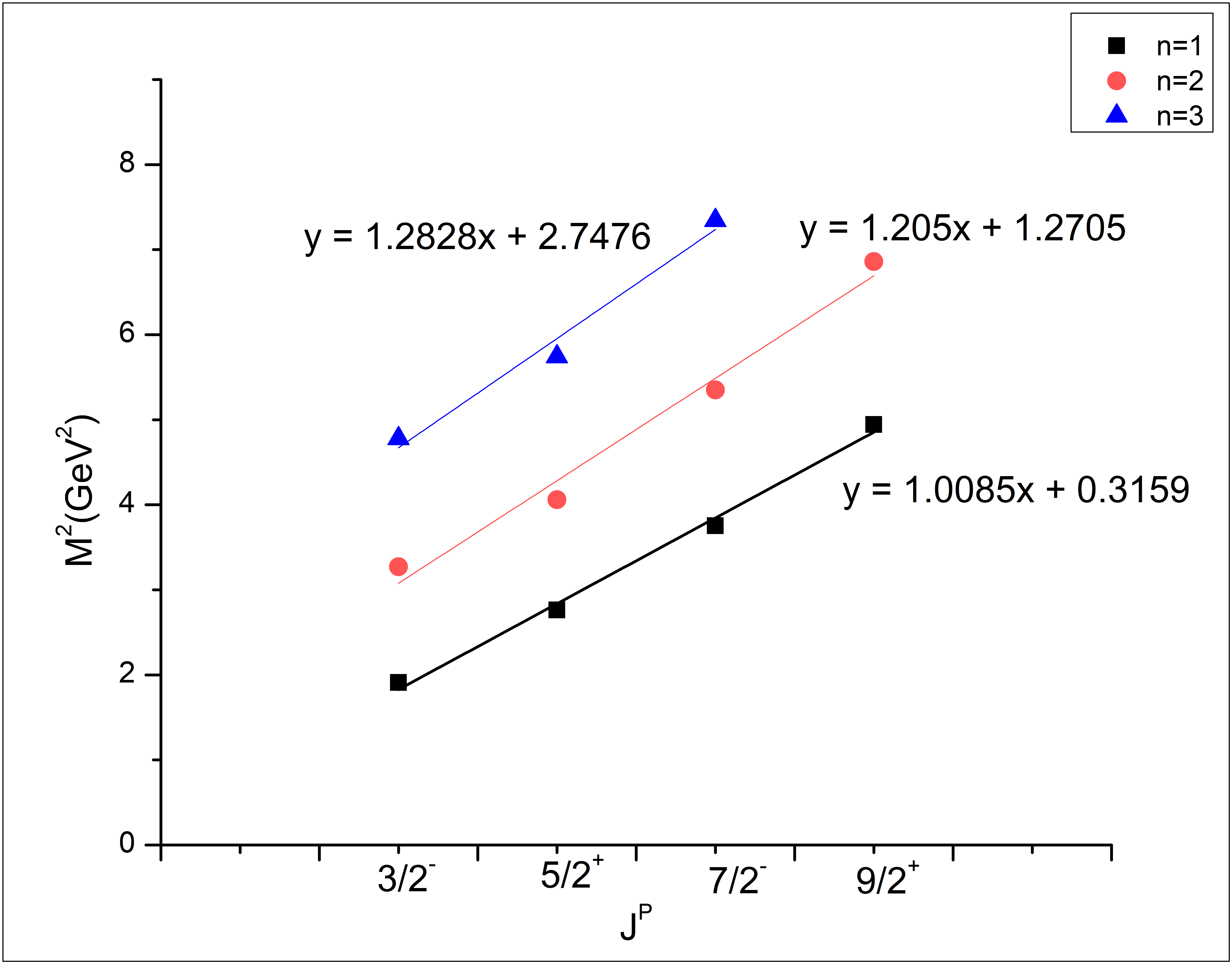}
	\caption{$J^P$ vs $M^{2}$ for $\Sigma$, Regge trajectory for Angular Momentum Quantum number J versus $M^{2}$ for unnatural parity.}
	\label{fig:sj2scr}
\end{figure}

A number of resonance masses to be fit for experimental comparison and observations, Regge trajectories play a significant role\cite{kan}.  Regge trajectories are basically plots of total angular momentum J and principle quantum number n against the square of resonance mass as depicted by figures \ref{fig:lscr}, \ref{fig:ljscr}, \ref{fig:sscr}, \ref{fig:sj1scr}, \ref{fig:sj2scr}. A correct spin-parity assignment for a state can perhaps be predicted using these plots \cite{guo}.
\begin{subequations}
	\begin{align}
		J = aM^{2} + a_{0} \\
		n = b M^{2} + b_{0}
	\end{align}
\end{subequations}
As it is evident that there are states in PDG whose spin-parity are not precisely known but resonance mass has been observed. Such points when put on the same line allows us to comment through the natural and unnatural parity positions. The value of slope and intercept allow us to extend it to further points to predict the higher excited states in any given baryon spectrum.
\section{Conclusion}
\label{sec5}
The present work summarizes the effect of screened type potential under hypercentral Constituent Quark Model (hCQM) for $\Lambda$ and $\Sigma$ baryons. So far, linear potential has been applied to light spectrum, whereas screened potential provided reasonable results for heavy quark systems \cite{review}. The screening parameter plays a role in determining the spin-split and mass at higher angular momentum states. The obtained masses have been comparable with the basis of experimental known values with different star status. The masses of higher spin state for a given L value, decreases in hierarchy. The hyperfine splitting is observed to be less with screened potential than those of linear one. 

\section*{Acknowledgments}
C. Menapara acknowledges the support from DST under INSPIRE Fellowship. Authors are thankful to organizers of ICNFP 2022.

\end{document}